\documentclass{aa}
\usepackage{graphicx}
\usepackage{txfonts}

\usepackage{natbib}

\usepackage{lscape}
\usepackage{longtable}

\begin{document}

\title{CNO and F abundances in the barium star
HD~123396}

\titlerunning{CNO and F abundances in the barium star
HD~123396}
\authorrunning{Alves-Brito et al.}

\author{
A.~Alves-Brito\inst{1,2} \and
A. I.~Karakas\inst{3} \and
D.~Yong\inst{3} \and
J.~Mel\'{e}ndez\inst{4} \and
S.~V\'asquez\inst{1} 
}


\institute{
Departamento de Astronom\'{\i}a y Astrof\'{\i}sica, Pontificia Universidad
Cat\'olica de Chile, Av. Vicu\~na Mackenna 4860, Santiago, Chile
\email{abrito@astro.puc.cl} 
\and
Centre for Astrophysics and Supercomputing, Swinburne University of
Technology, Hawthorn, Victoria 3122, Australia 
\and
Research School of Astronomy and Astrophysics,
The Australian National University, Cotter Road, Weston, ACT 2611, Australia
\and
Universidade de S\~{a}o Paulo, IAG, Rua do Mat\~{a}o 1226, 
Cidade Universit\'{a}ria, S\~{a}o Paulo 05508-900, Brazil.
}

\date{Received: ; accepted: }

\abstract
{Barium stars are moderately rare, chemically peculiar objects, which are
believed to be the result of the pollution of an otherwise normal star by material from an evolved companion on the asymptotic
giant branch (AGB).
}
{We aim to derive carbon, nitrogen, oxygen, and fluorine abundances 
for the first time from the infrared spectra of the barium red giant star HD 123396
to quantitatively test AGB nucleosynthesis models for producing barium
stars via mass accretion.}
{High-resolution and high S/N infrared spectra were obtained using the
Phoenix spectrograph mounted at the Gemini South telescope.
The abundances were obtained through spectrum synthesis of individual
atomic and molecular lines,
using the MOOG stellar line analysis program, together with Kurucz's stellar
atmosphere models. The analysis was classical, using 1D stellar models and
spectral synthesis under the assumption of local
thermodynamic equilibrium.
}
{We confirm that HD 123396 is a metal-deficient barium star ([Fe/H] = $-$1.05),
with $A(\mathrm{C}) = 7.88$, $A(\mathrm{N}) = 6.65$, $A(\mathrm{O}) = 7.93$, and
$A(\mathrm{Na}) = 5.28$ on a logarithmic scale where $A(\mathrm{H}) = 12$, leading to
[(C+N)/Fe] $\approx$ 0.5. The $A(\mathrm{CNO})$ group, 
as well as the $A(\mathrm{Na})$ abundances, is in excellent agreement with those
previously derived for this star using high-resolution optical data. 
We also found $A(\mathrm{F}) = 4.16$, which implies [F/O] = 0.39, a
value that is substantially higher than the F abundances measured
in globular clusters of a similar metallicity, noting that there are
no F measurements in field stars of comparable metallicity.
}
{The observed abundance pattern of the light elements (CNO, F, and Na)
recovered here as well as the heavy elements ($s$-process) studied elsewhere suggest that the surface composition of HD 123396
is well fitted by the predicted abundance pattern of a 1.5{\it M$_{\odot}$}
AGB model star with Z = 0.001. Thus, the AGB mass transfer hypothesis 
offers a quantitatively viable framework.}


\maketitle
%

\section{Introduction}

Barium stars are chemically peculiar objects --- dwarfs, subgiants, and giants
--- that show an excess of carbon and s-process elements in their
atmosphere \citep[e.g.][]{tomkin89,barbuy92,allen06a,smilj07}.
Identification of stars showing strong enhancements of neutron
capture elements that can be produced by the $slow$ neutron capture
process (the $s$-process), as well as prominent CH, CN, and C$_{\rm 2}$ 
molecular lines, dates back to \citet{bidelman51}. Since then,
many other authors have sought to measure abundances of different
species in order to better constrain the possible mechanisms by which barium stars can be formed. 

The first detailed analysis of the barium star HD 46407 was by
\citet{burbidge57}, who found overabundances of barium and 
many other elements relative to solar or normal giants. 
Carbon, nitrogen, and oxygen were first analysed in barium stars
by \citet[][HR 774]{tomkin79} and \citet[][$\zeta$ Cap]{smith84}, 
who demonstrated that 
C were roughly solar and O slightly deficient by 0.1
dex. Nitrogen, on the other hand, was found to be
enhanced by 0.3 dex with respect to the Sun in the barium stars analysed. 
More recently, \citet{allen06a} and \citet{allen07} acquired 
high-resolution spectra of giant and dwarf barium
stars and obtained chemical abundances for 
several heavy elements(Sr, Y, Zr, Ru, Ba, La, Ce, Pr, Nd, Sm, Hf, Pb), that are 
produced by neutron-capture nucleosynthesis 
\citep[the $s$ and $r$ processes, see e.g.,][]{sneden08}.
In summary, detailed abundance analyses of the barium stars 
reveal that the heavy 
elements produced by the $s$-process are enhanced by factors of 
2 -- 30 with respect to normal giants, whereas C is overabundant 
by a factor of 3 in the most extreme cases. While barium and CH stars
likely had mass transferred from a former AGB companion, the 
latter group show strong C$_{2}$ molecular bands and higher 
C/O abundance ratios when compared to the former
\citep[e.g.,][]{luck91,mcclure97b}. The general view is that CH 
stars are more metal poor than barium stars, though there is no clear 
boundary.

The $slow$ neutron capture process occurs in the advanced 
thermally-pulsing asymptotic giant branch (TP-AGB) phase of low and 
intermediate-mass stars \citep[e.g.,][]{busso99}. For this reason 
significant enrichment of $s$-process elements are not expected in stars 
in earlier evolutionary phases. The discovery that all observed barium
stars belong to spectroscopic binary systems 
\citep{bohm80,mcclure80,mcclure83,mcclure84,bohm85,jorissen88,mcclure90,udry98a,udry98b}
with eccentricities significantly lower than a sample of spectroscopically 
normal G and K giants \citep{mcclure90,jorissen98} is evidence that the C 
and large $s$-process overabundances are the result of accretion of 
enriched material from a former TP-AGB companion rather than the result of 
internal nucleosynthesis. The accretion of material is the direct result 
of mass transfer via Roche Lobe overflow or stellar winds,
and the low eccentricities arise from tidal circularization 
\citep{boffin88,han95,karakas00,izzard10}. In this context, CNO elements, 
along with $s$-process elements, are key tracers of the nucleosynthesis in the
previous AGB progenitor, so they provide crucial constraints on the
formation mechanism. According to the mass-transfer hypothesis, the former AGB stars would now 
be optically invisible white dwarfs, and several investigations with the 
IUE satellite have found the characteristic UV excesses typical of white 
dwarfs \citep[e.g.][]{bohm00a,bohm00b}.

Another unique tracer of AGB nucleosynthesis is the element F. Outside of the
solar system, F abundances were
first determined in the pioneering work of \citet{jorissen92} who
analysed the correlation between F and C abundances in a sample of 
AGB stars. While it was clear from that study that F can be
synthesized in AGB stars, and it is produced in the He-intershell 
along with C and $s$-process elements \citep{jorissen92,lugaro04}, 
the nucleosynthetic origin of F in the Galaxy is still 
somewhat of a mystery. 
To date, F abundances have been determined for giant stars in the 
Galactic and extragalactic field 
\citep[e.g.,][]{jorissen92,cunha05,pandey06,schuler07,pandey08,cunha08,utt08,abia10}, 
in Galactic and extragalactic globular clusters
\citep{cunha03,smith05,yong08c,lebzelter08},
and in post-AGB stars and planetary nebulae \citep{werner05,zhang05,otsuka08}.
For barium stars in particular, F abundances have been measured 
for only two stars \citep{jorissen92} and thus much work
is required on both the theoretical and observational sides to improve not only
our knowledge of the nature of barium stars but also the astrophysical
environment under which F is synthesized.

HD~123396 is of particular interest because it has an iron abundance
of [Fe/H]\footnote{We have adopted the usual spectroscopic notation, where {\it A(X) = log[n(X)/n(H)]
+12} and {\it [X/Y] = log[n(X)/n(Y)]$_{*}$ $-$ log[n(X)/n(Y)]$_{\odot}$}, where {\it n}(X) and
{\it n}(Y) are the number particle density of "X" and "Y" respectively.} 
$\approx -1$, and is the most metal-deficient barium star from
the \citet{allen06a} sample. The low metallicity of HD~123396
makes it unique among barium stars as it is right at the low end of the
barium star metallicity distribution, and at a metallicity that
encompasses the metal-rich tail of the halo, as well as the metal-poor 
tail of the disk \citep[e.g.,][]{luck91,junq01,allen06a}. 

The binary nature of HD 123396 has not been definitively established. 
There is no evidence, as yet, for radial velocity variations. 
\citet{jorissen05} employed an additional check for binarity involving 
the comparison between the Hipparcos and Tycho-2 proper motions. However, this
technique has been successfully employed for stars with parallaxes higher than
5 mas, while HD 123396 has a parallax of 1.78 $\pm$ 0.73 mas \citep{vl07}.

Here in this study we present the first abundance estimates of
C, N, O, and F for HD~123396 using spectra taken at infrared
wavelengths using the high-resolution, near-infrared Phoenix spectrograph
at Gemini-South. HD~123396 exhibits overabundances of heavy elements 
at the three $s$-element peaks: the light $s$-peak, which includes the
elements Sr, Y, and Zr, and denoted by the ratio [ls/Fe]; the heavy
$s$-peak, which includes Ba, Ce, Nd, and Sm, and is denoted
by the ratio [hs/Fe]; and the ratio of [Pb/Fe]. This makes HD~123396
an excellent tracer of chemically peculiar objects at low metallicities.
The abundances of the CNO elements and F need to be 
considered together with the $s$-process elements, because in the
context of the AGB mass transfer hypothesis, C and F are produced 
together with the heavy elements. 

\section{Observations}

We acquired high-resolution and high signal-to-noise (S/N) spectra of HD
123396 ($\alpha$ = 14$^{\rm h}$17$^{\rm m}$33$^{\rm s}$.239, $\delta$ =
$-$83$^{\rm o}$32$^{\rm m}$46$^{\rm s}$.67 : J2000)
using the spectrograph Phoenix \citep{hinkle03} on the
Gemini-South 8m telescope at Cerro Pachon, Chile, in 2009 June, under project
GS-2009A-Q-26. The star was selected from the
high-resolution optical abundance analysis carried out by
\citet{allen06a}. 

HD~123396 was observed in service mode using basically the same setup described
in \citet{yong08c}, which allowed us to cover 1551 $\leq$ $\lambda$ $\leq$ 1558 nm
in the H band and 2330 $\leq$ $\lambda$ $\leq$ 2340 nm in the K band, both employing a slit
width-size of 0.34 arcseconds in order to reach a 
spectral 
resolution of R = 50,000. Four
exposures of HD~123396 of 51s each were taken at two different positions along
the slit in the H band (H = 6.3 mag) and of 80s each in the K band (K$_{\rm s}$
= 6.1 mag). 
The individual spectra in each band were co-added leading to S/N = 200
in each band. The spectra were reduced with IRAF\footnote{IRAF is distributed by the National
Optical Astronomy Observatory, which is operated by the Association of
Universities for Research in Astronomy (AURA) under cooperative agreement with
the National Science Foundation.} employing standard procedures, which included
dark and sky subtraction, flatfield correction, spectrum extraction, 
wavelength calibration, and telluric correction. 

\section{Analysis}

To calculate reliable chemical abundances, 
it is necessary to obtain precise stellar
atmospheric parameters such as temperature ($T_{\mathrm{eff}}$), surface gravity
($\log g$), microturbulent velocity ($\xi$) and metallicity
([Fe/H]). We assumed a typical mass of 0.8{\it M$_{\odot}$} for the star
HD 123396 \citep[e.g.,][]{vand00}.
Using optical data, \citet{allen06a} obtained photometric and spectroscopic
parameters for this star. While the photometric parameters were
obtained from optical and infrared colours by employing
different temperature calibrations \citep[e.g.,][]{alonso99,ramirez04} and
classical equations of stellar evolution, the spectroscopic parameters were
calculated by imposing excitation (temperature) and ionization (gravity)
equilibrium of $\ion{Fe}{i}$ and $\ion{Fe}{ii}$ lines. Here we have
adopted the values given in  \citet{allen06a}  as our first guess. However,
using our own infrared data, the final atmospheric parameters (basically
$T_{\mathrm{eff}}$
and [Fe/H]) were iteratively adjusted (see Table \ref{t:atmos}).

Our abundance calculations were carried out using 
Kurucz model atmospheres \citep{castelli97}, the local thermodynamic equilibrium (LTE) spectral
synthesis program MOOG \citep{sneden73}, and the same line list presented and
described in previous studies \citep[e.g.,][for more details]{melendez03,melendez08,yong08c}.

For consistency, we have checked the line list in both H and K bands by fitting
the Arcturus spectrum \citep[][$R=100,000$]{hinkle95} with CNO abundances
from \citet{ryde10} and with F and Na abundances as given in
\citet{cunha03}. A macroturbulent velocity of 3.7 km~s$^{-1}$ was
adopted \citep{ryde10}. For HD 123396, we then derived chemical
abundances of C, N, O, F, Na, and Fe using OH,
CO, and CN molecular bands and $\ion{Fe}{i}$ lines in the H band at 1555 nm, as well as
the HF (1-0) R9 line at 2335 nm ($\chi$ = 0.480 eV, $\log gf$ = --3.955
dex) and the NaI line at 2337 nm in the K band. We note that the CO
bandhead absorption at 1558 nm was discarded from the analysis. 
As seen in Fig. \ref{f:synth}, the normalization at the
extreme red edge of the spectrum makes the abundance fit very uncertain. Had
we adjusted the CO molecular lines in the H band, the [C/Fe]
inferred from the CO lines in the H and K bands would be different from the CO
lines in the K band yielding systematically higher abundances of C by 0.40 dex.
To compute the synthetic spectra, we initially used the abundances
derived by \citet{allen06a} and then simultaneously changed C, N, and O
abundances to obtain the best fits.

\begin{figure}
   \centering
   \includegraphics[width=9.5cm]{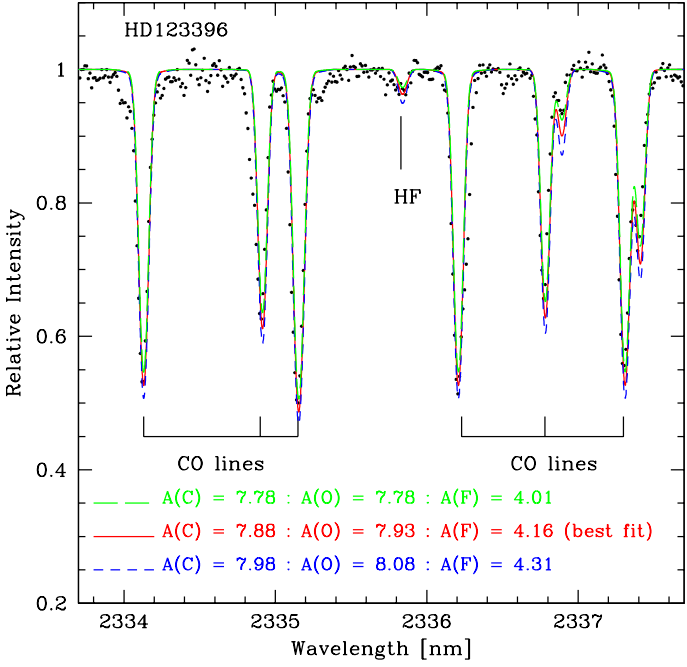}
   \caption{Observed ({\it
   points}) and synthetic ({\it line}) spectra of HD~123396 for a sample of the
   wavelength regions in the K band. The abundances derived for C, O, and F are
   labelled in the figure. The synthetic spectra stand for the best fit ({\it
   solid line}) and unsatisfactory ones ({\it short and long dash lines}) by
   varying the abundances of C by $\pm$0.10 dex, and O and F by $\pm$0.15 dex,
   respectively (refer to the text).}
              \label{f:synth_fluor}%
    \end{figure}

To take the instrumental profile and other broadening effects (e.g.
macroturbulence) into account, the raw spectra were convolved with Gaussian
functions of FWHM $\approx$ 0.40$\rm \AA$ in the H band and FWHM $\approx$ 0.60$\rm
\AA$ in the K band. In Figs. \ref{f:synth} and \ref{f:synth_fluor} we show an example of spectral
synthesis in the different bands.

\section{Results and discussion}

\begin{figure*}
   \centering
   \includegraphics[width=13cm]{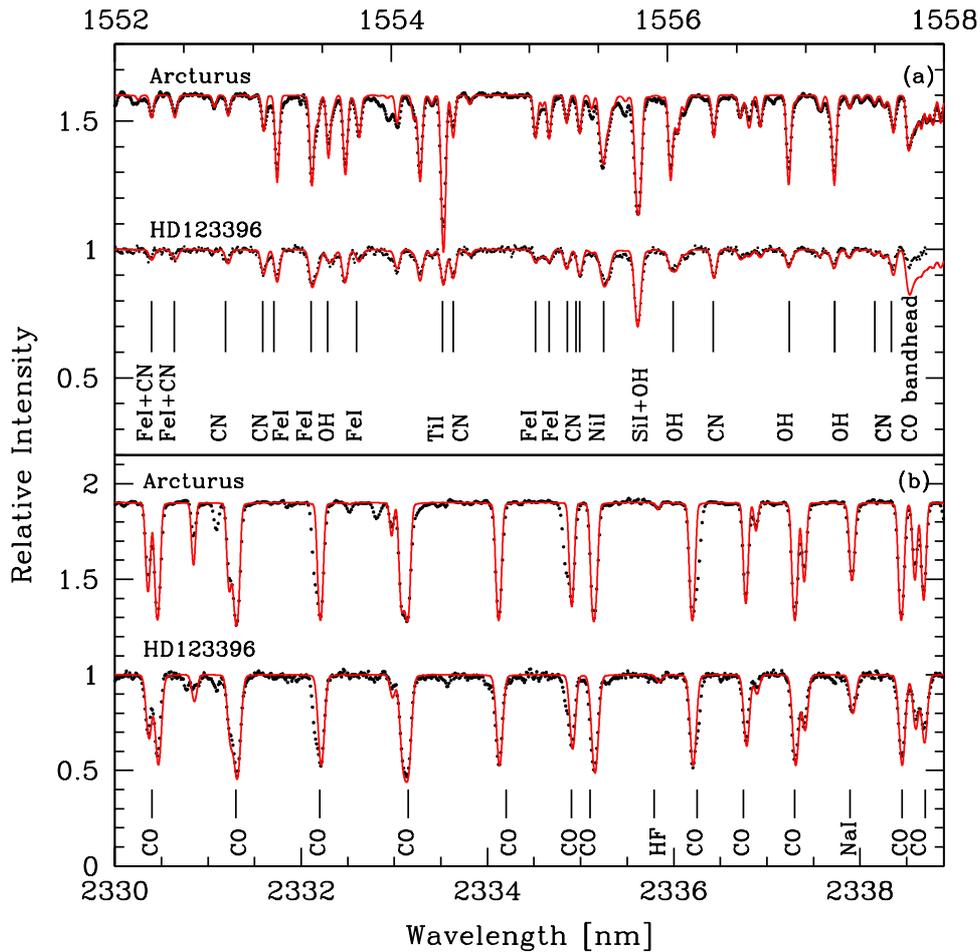}
   \caption{Observed ({\it
   points}) and synthetic ({\it lines}) spectra of Arcturus and HD~123396 in
   the H {\it
   (a)} and K {\it (b)} bands. 
    Several atomic and molecular lines are identified. Refer to the text
   for more details.}
              \label{f:synth}%
    \end{figure*}

In Tables \ref{t:atmos} and \ref{t:abundances} we present the final atmospheric
parameters and abundances derived for the star HD~123396. We
also compare our results with those obtained by \citet{allen06a} for the
same star and for the two barium stars with F abundances previously derived
by \citet{jorissen92}.
The primary source of errors in the abundances are the 
errors in the atmospheric parameters. Varying the atmospheric
parameters by their typical uncertainties ($\Delta$$T_{\mathrm{eff}}$ = $\pm$ 100~K, $\Delta$$\log g$ = $\pm$ 0.3~dex, 
$\Delta\xi$ = $\pm$ 0.2 kms$^{-1}$ and $\Delta$[Fe/H] = $\pm$0.1 dex),
we find that C, N, and O are uncertain by 0.10, 0.12, and 0.15 dex when
all errors are added in quadrature. The Na
and F abundances change by 0.05 dex and 0.15 dex, respectively. 
These values suggest that the total error in the [F/O] ratio cancels out.
In Figure \ref{f:synth} we display an example of the spectral synthesis
performed for the star in the H and K bands compared with that of Arcturus.

\setlength\tabcolsep{2.5pt}
\begin{table}
\begin{flushleft}
\caption{Stellar parameters for HD~123396 and two barium stars from the
literature with previous F abundance determinations.}
\label{t:atmos}      
\centering          
\begin{tabular}{lccccccc}     
\noalign{\smallskip}
\hline\hline    
\noalign{\smallskip}
\noalign{\vskip 0.1cm} 

Star &  $T_{\mathrm {eff}}$ & $\log g$ & $\xi$ & v$_{\rm r}$ & [$\ion{Fe}{i}$/H] & [$\ion{Fe}{ii}$/H]& Ref.\\ 
     &  [K] & [dex] & [kms$^{-1}$] & [kms$^{-1}$] & [dex] & [dex] & \\   
\noalign{\vskip 0.1cm} 
\hline 
\noalign{\vskip 0.1cm} 
HD~123396 &  4400 & 1.40 & 1.20 & 26.40 &  $-$1.05 &  ...     & (1) \\  
          &  4360 & 1.40 & 1.20 & 26.64 &  $-$1.19 &  $-$0.99 & (2) \\
	  &  4480 & 1.20 & 1.20 & 26.64 &  $-$1.19 &  $-$0.99 & (2) \\
HD~121447 &  4200 & 0.80 & 2.50 & ...   &  $+$0.05 &  ...     & (3) \\	  
HD~178717 &  4300 & 1.00 & 2.20 & ...   &  $-$0.18 &  ...     & (3) \\		  
\hline 
\end{tabular}
\begin{minipage}{.88\hsize}
 Notes.--- (1): This work; (2): \citet{allen06a}; (3): \citet{smith84}.\\
\end{minipage}			
\end{flushleft}
\end{table}  

\setlength\tabcolsep{2.5pt}
\begin{table}
\begin{flushleft}
\caption{Abundances for the star HD~123396 and two barium stars from the
literature with previous F abundance determinations.}
\label{t:abundances}      
\centering          
\begin{tabular}{lccccccc}     
\noalign{\smallskip}
\hline\hline    
\noalign{\smallskip}
\noalign{\vskip 0.1cm} 

Star & $A(\mathrm{C})$ & $A(\mathrm{N})$ & $A(\mathrm{O})$ & $A(\mathrm{F})$ &
$A(\mathrm{Na})$ & [F/O] &  Ref. \\  	       
\noalign{\vskip 0.1cm} 
\hline 
\noalign{\vskip 0.1cm} 
HD~123396 & 7.88 & 6.65 & 7.93  &  4.16 & 5.28 & $+$0.39 & (1) \\  
          & 7.83 & 6.68 & 7.97  &  ...  & 5.26 & ...	 & (2) \\   
HD~121447 & 8.70 & 8.40 & 8.80  &  5.35 & ...  & $+$0.71 & (3,4) \\   
HD~178717 & 8.60 & 8.50 & 8.70  &  5.24 & ...  & $+$0.70 & (3,4) \\   
\hline 
\end{tabular}
\begin{minipage}{.88\hsize}
 Notes.--- (1): This work; (2): \citet{allen06a}; (3, 4):
 \citet{smith84,jorissen92}. 
 In this work, $A(\mathrm{C,N,O})_{\odot}$ = $8.42$, $7.82$, and $8.72$;
 $A(\mathrm{F,Na,Fe})_{\odot}$ = $4.56$, $6.17$ and $7.50$ \citep[see][]{asplund05,melendez08}.
The [F/O] ratios from Ref. (4) were renormalized to our adopted solar abundances. \\
\end{minipage}			
\end{flushleft}
\end{table}  

Using high-resolution optical spectra, \citet{allen06a} performed a detailed
abundance analysis of HD~123396, yielding chemical abundances for C, N, $\alpha$,
iron-peak, and neutron-capture elements. 
As can be seen in Table \ref{t:atmos}, we obtain [Fe/H] =
$-$1.05 for this star, which is in excellent 
agreement with the mean value found
by \citet{allen06a}, that is, [Fe/H] = $-$1.09.
For the four other elements in common 
(see Table \ref{t:abundances}), we
have found $\Delta$$A(\mathrm{C,N,O}) = $+$0.05$ dex, $-$0.03 dex and $-$0.04
dex, but $\Delta$$A(\mathrm{Na}) = +0.02$ dex. 
While the differences in the absolute abundances are statistically
insignificant, we recall that the CNO group elements are better
studied in the infrared, where their lines are more numerous, and the
continuum is formed deepest in the layers of the stellar atmosphere due to the
opacity minimum of H$^{-}$ near the H band.
As presented in Table 7 of \citet{barbuy92}, the C abundances can
vary up to +0.25 dex for the same star depending on the optical molecular lines used (C$_{\rm
2}$ or CH lines), likely from 3D and NLTE effects.

Using high-resolution infrared spectra acquired
with the Fourier transform spectrometer on the Kitt Peak National
Observatory's 4~m telescope, \citet{jorissen92} were the first to obtain
F abundances for a large and heterogeneous sample of giant stars, which
included two metal-rich barium stars.
From Table \ref{t:abundances}, one can see that HD~123396 has an [F/O]
ratio that is $\sim$ +0.30 dex lower than the two barium stars analysed 
by \citet{jorissen92}. This result may not be surprising for
the following reasons. First, the metallicity of HD 123396 is lower
by roughly an order of magnitude than the other two barium 
stars \citep[based on the iron abundances quoted by ][]{smith84}.
Second, the recent analysis by \citet{abia10} resulted in 
significantly lower F abundances than in previous studies in the
literature. The lower F abundances are associated with improved model
atmospheres and atomic and molecular line lists in the 2.3 
$\mu$m region, and this has resulted in the identification of molecular
blends. 

\citet{allen06a} show that HD~123396 is an $s$-process rich 
star (e.g., [Pb/Fe] = 1.00). Given that low and intermediate-mass
stars produce $s$-process elements during the AGB phase, and that 
barium stars are, generally, less evolved and {\em not} on the AGB
leads to the conclusion that they obtained their heavy-element 
overabundances via mass transfer from a previous AGB companion. 
The high percentage of barium stars in wide binary systems 
is another piece of evidence for this hypothesis. Further evidence 
comes from the relatively high F abundance of HD~123396 that we 
derive in this work.

While the astrophysical site for F production in the Galaxy is
uncertain with massive stars perhaps playing a dominant role
\citep{meynet00,cunha03,renda04}, there is strong evidence 
that F is produced in AGB stars \citep{jorissen92,abia10}. 
Possible reaction pathways for F production 
in AGB stars were outlined by \citet{jorissen92} and examined in
detail by \citet{forestini92}, \citet{mowlavi96}, \citet{lugaro04}, and
\citet{karakas08}. The F production in AGB stars occurs in the
He-intershell via a complex series of proton, $\alpha$, and 
neutron-capture reactions, starting with the
$^{14}$N($\alpha$,$\gamma$)$^{18}$F($\beta$$^{+}$)$^{18}$O({\it
p},$\alpha$)$^{15}$N($\alpha$,$\gamma$)$^{19}$F reaction. The protons for the CNO cycle reaction $^{18}$O($p,\alpha$)$^{15}$N
come from the $^{14}$N($n,p$)$^{14}$C reaction, which in turn requires 
free neutrons. $^{14}$N is one of the the main neutron 
poisons in AGB stellar models \citep[see discussion in][]{busso99}.  
The $^{13}$C($\alpha$,n)O$^{16}$ reaction is the dominant 
source of free neutrons in low-mass AGB stars and is the main
neutron source for the operation of the $s$ process. 
The $^{22}$Ne($\alpha$,n)Mg$^{25}$ neutron source also produces a small
burst of neutrons during convective thermal pulses but models
and observational evidence suggests that it plays a minor role
\citep{busso01,abia02}.  Fluorine can also be destroyed via He-shell
burning via $^{19}$F($\alpha,p$)$^{22}$Ne at temperatures over 
about 300$\times 10^{6}$K, and  F production has been shown to be
a sensitive function of the initial mass and metallicity, with
peak production occurring at $\approx 3 {\it M_{\odot}}$ at $Z=0.02$, which
is reduced to $\approx 2 {\it M_{\odot}}$ at $Z=0.0001$ ([Fe/H] $=-2.3$)
\citep{lugaro04,karakas10a}. Furthermore, because F production occurs in the He-intershell, it is dredged to the
surface via the repeated action of the third dredge-up, along
with C and any $s$-process elements. 

\begin{figure}
   \centering
   \includegraphics[width=9.5cm]{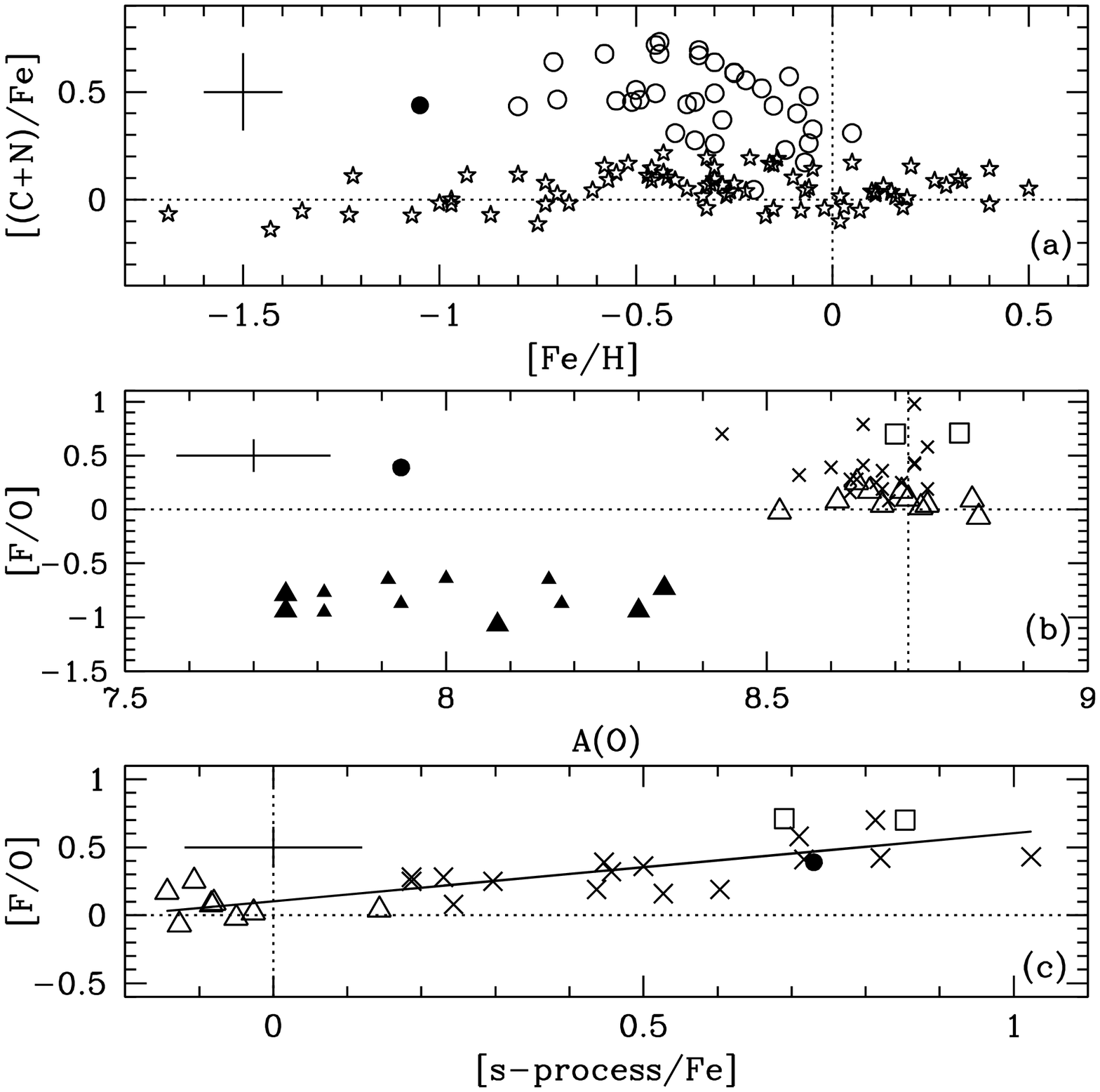}
   \caption{{\it (a)}: [(C$+$N)/Fe] vs. [Fe/H] for normal K giants ({\it stars}:
   \citet{melendez08,ryde10}) and barium stars in this work ({\it
filled circle}) and from the literature 
\citep[{\it open circles}:][]{smith84,barbuy92,allen06a}. ({\it b}): [F/O] vs. $A(\mathrm{O})$ for HD~123396 ({\it filled circle})
compared to (i) field stars --- two barium stars 
\citep[{\it open squares}:][]{jorissen92}, M giant stars
\citep[{\it open triangles}:][]{cunha03}, MS and S stars
\citep[{\it crosses}:][]{jorissen92}; and compared to (ii) GC's stars --- M4 \citep[{\it small filled triangles}:][]{smith05} and NGC 6712 
\citep[{\it large filled triangles}:][]{yong08c}.
All abundances were renormalized to $A(\mathrm{F})_{\odot}$ = $4.56$ and
$A(\mathrm{O})_{\odot}$ = $8.72$. ({\it c}): [F/O] vs. [s-process/Fe], where
s-process stand for Y, Zr and Nd. The s-process abundances were 
taken from \citet[][]{allen06a}, 
\citet[][{\it two barium stars}]{smith84}, and 
\citet[][{\it M, MS and S stars}]{smith85,smith86b,smith90a}.
The dotted lines indicate the solar abundances. 
Typical uncertainties are quoted.}
              \label{f:abund}%
    \end{figure}

In Fig. \ref{f:abund}a we show a plot of [(C+N)/Fe] vs. [Fe/H] for normal
giant stars from the literature \citep{melendez08,ryde10} compared with
those obtained for barium stars in this work and from the literature
\citep{smith84,barbuy92,allen06a}. This figure suggests that the
[(C+N)/Fe] ratio we recover for HD~123396 lies
on, or near, a linear extrapolation based on the more metal-rich barium stars, that
is, [(C+N)/Fe] $\approx$ 0.5. On the other hand, {\it normal} K giant stars
that have not experienced the first dredge up yet have [(C+N)/Fe]
$\approx$ 0. We point out that some of the stars that were previously classified as the most
metal-poor halo barium stars are now shown in fact to be CH or AGB
stars \citep[see e.g.,][]{jorissen05,drake08,pereira09}. However,
barium stars present longer periods and greater eccentricities than the CH
stars \citep[e.g.][]{vanture92a}.
From a chemical point of view, while classical barium and CH stars are
$s$-process enriched to a similar degree, CH stars are differentiated 
by their strong C$_{\rm 2}$ molecular bands, high C/O
abundances ([C/O] $\geq$ 1) and [(C+N)/Fe] $\approx$ +1.0
\citep[e.g.][]{vanture92a,vanture92b}.

In Fig. \ref{f:abund}b we plot [F/O] vs. $A(\mathrm{O})$  for the star
analysed in this work compared with other two disk barium stars, cool K,
M, MS, and S giant stars from \citet{jorissen92} and \citet{cunha03}, as
well as stars from the GCs M4 and NGC 6712
\citep{smith05,yong08c}, which have a metallicity similar
to the barium star analysed in this work. For HD~123396, 
we find [F/O] = +0.39 dex, which means that, for
the three barium stars analysed to date, $\langle$[F/O]$\rangle$ = +0.60 $\pm$
0.18 dex. Interestingly, \citet{smith05} and \citet{yong08c}
report, respectively,
$\langle$[F/O]$\rangle$ = $-$0.77 $\pm$ 0.13 dex (N = 7 stars) and
$\langle$[F/O]$\rangle$ = $-$0.89 $\pm$ 0.13 dex (N = 5 stars) for the GCs M4 ([Fe/H] = $-$1.18) and NGC 6712 ([Fe/H] = $-$0.96). 
Our measurement of F in HD 123396 is 
the first measurement of F in a star whose metallicity, $A(\mathrm{O})$, 
overlaps with the GC data. From Fig. \ref{f:abund}b one can infer that the F/O
ratio in HD~123396 is significantly different from the mean values found in the
GCs studied, $\delta$(F/O) 
$\simeq$ 1.2. This difference corresponds to $\sim$10-$\sigma$ or a factor of
$\sim$16 higher than the uncertainty estimated above for the [F/O] ratio of HD~123396.
In addition, the GC stars and HD~123396 are all metal-poor giants such that
dwarf/giant systematic errors cannot be invoked to explain the abundance
difference. 
However, HD 123396 is a chemically peculiar barium star that probably 
received its F from AGB mass transfer. According to this hypothesis, barium stars
will likely have enhanced F abundances relative to {\it normal} field stars of
the same metallicity. Thus, additional F measurements in metal-poor field stars are
necessary  to test if there is a real discrepancy in F abundances measured in 
{\it normal} field stars and GCs at such metallicity.

In Fig. \ref{f:abund}c we present the [F/O] abundances as a function 
of the mean abundances of two {\it light} $s$-process elements from 
the first $s$-element peak (Y and Sr) and one from the 
second $s$-process peak (Nd). The abundances were taken from
different sources \citep{smith85,smith86b,smith90a,allen06a}, and
we note that no attempt was made to take likely systematic effects into
account. Clearly, HD~123396 follows the general trend for [F/O] to increase with
$s$-process elements as found by \citet{jorissen92}. From a linear fit to
the data it turns out that the
scatter in the data is $\approx$ 0.15 dex, which is likely due to 
the convolution of measurement uncertainties from individual studies 
and the errors introduced by an inhomogeneous sample.

\begin{figure}
   \centering
   \includegraphics[width=9.5cm]{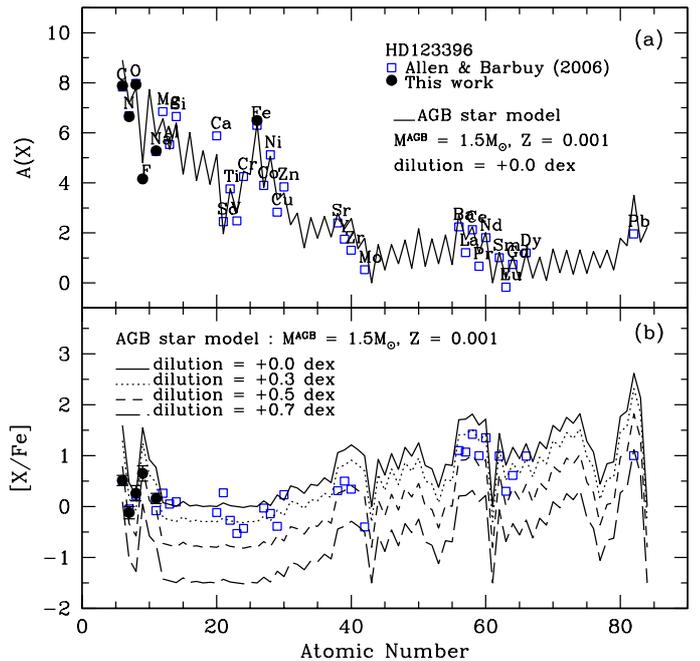}
   \caption{The barium giant star HD~123396 observed abundance pattern compared
   with theoretical predictions of a 1.5{\it $M_{\odot}$} AGB star of $Z=0.001$. 
   Symbols and details on
   the model used are given in the figure and text.}
              \label{f:pattern}%
    \end{figure}

In Fig. \ref{f:pattern} we show the abundances derived
for HD~123396 in this work and in \citet{allen06a}, along with
nucleosynthesis predictions from a
1.5{\it M$_{\odot}$}, $Z$ = 0.001 AGB star model. The AGB model was
calculated using the stellar evolution code described in detail
by \citet[][and references therein]{karakas10a}, which uses 
the \citet{vw93} mass-loss rate on the AGB, and with the addition 
of C and N-rich low-temperature opacities tables from \citet{marigo09}.
The $s$-process abundance predictions were calculated using the 
nucleosynthesis code described in \citet{karakas10a} and 
\citet{karakas10c}, updated to use a full network of 320 species 
for elements from hydrogen through to bismuth, reaction 
rates taken from the JINA REACLIB library \citep{cyburt10}.
Solar abundances of C, N, O, Ne, Mg, Si, S, Ar, and Fe are 
the pre-solar nebula values from
Table 5 of \citet{asplund09}; F is the meteoritic value of 
$A(\mathrm{F})_{\odot}$ = $4.42$ from Table 1 of the same
paper (chosen because it has a lower uncertainty), and for
many of the elements heavier than Fe we use the meteoritic values 
for the solar abundances (e.g., Ga, Sr, Eu, Pb).
In order to obtain
an enrichment of $s$-process elements, an exponentially
decaying abundance of protons is artificially introduced into 
the top layers of the He-intershell
at the deepest extent of each third dredge-up episode. 
We do this in exactly the same manner as outlined in 
\citet{lugaro04} and \citet{karakas10a}. This {\em partially
mixed zone} is required in order
to facilitate the formation of a so-called $^{13}$C pocket, which allows neutrons to be released via the
$^{13}$C($\alpha$,n)O$^{16}$ reaction \citep[see][for more
details]{straniero95,gallino98,gori00,herwig05}. The mass of the proton
profile is a free parameter that we set as a constant mass. 
We adopt two choices for the extent of the partially mixed
proton zone: 1) 0.002{\it $M_{\odot}$} and 2) 0.0004{\it $M_{\odot}$} (or a 
fifth of the mass of our {\em standard} choice).
We discuss the implications of this choice below.
In Table~\ref{t:agbmodel} we present the surface composition
for a selection of elemental abundances for the two choices 
of the partially mixed zone. These abundances from the tip of the
AGB and represent the {\em undiluted AGB composition}.

\begin{table*}
\centering
\caption{Predicted surface composition from a scaled-solar 
1.5{\it M$_{\odot}$}, $Z$ = 0.001 star model at the tip of the AGB. 
The [Fe/H] of the model star is $-1.18$, and initial abundances
are given in the first row.}
\label{t:agbmodel}      
\centering          
\begin{tabular}{lcccccccccc}     
\noalign{\smallskip}
\hline\hline    


Pocket & $A(\mathrm{C})$ & $A(\mathrm{N})$ & $A(\mathrm{O})$ & $A(\mathrm{F})$ & [F/O] & [Y/Fe] & [Zr/Fe] &
[Ba/Fe] & [Nd/Fe] & [Pb/Fe] \\  	       
\noalign{\vskip 0.1cm} 
\hline 
\noalign{\vskip 0.1cm} 
Initial & 7.29 & 6.69 & 7.55 & 3.24 & 0.00 & 0.00 & 0.00 & 0.00 & 0.00
& 0.00 \\ \hline
\noalign{\vskip 0.1cm}
0.002   & 8.91 & 7.17 & 7.79 & 4.81 & 1.32 & 1.11 & 1.21 & 1.70 & 1.71 & 2.61 \\
0.0004  & 8.94 & 7.17 & 7.74 & 4.59 & 1.16 & 0.51 & 0.60 & 1.05 & 1.07 & 1.96 \\
\hline 
\end{tabular}
\begin{minipage}{.88\hsize}
 Notes.--- The [F/O] and [X/Fe] abundances ratios were normalized to the 
solar abundances from \citet{asplund09}. \\
\end{minipage}			
\end{table*}  

The F abundance of the model star at the tip of the 
AGB is between [F/O] = 1.16--1.32 (depending on the mass of the 
partially-mixed zone assumed in the model, see
Table~\ref{t:agbmodel}). Using the dilution formula from 
\citet[][their equation 6]{bisterzo10} we can estimate the amount
by which the AGB surface composition has been diluted into the 
observed star's envelope. Larger dilution factors can indicate 
that only a small amount of material was transferred, and/or that
the star is now a giant with a large convective envelope.
To match the measured F abundance of +0.60 $\pm$
0.18, we need to dilute the predicted [F/O] ratio of 
1.16--1.32 (see Table~\ref{t:agbmodel}) by 0.5--0.7
dex, depending on the size of the partially mixed zone.
We note that our simple approach does not take into account the
fact both O and F vary between predictions and observations, and is
used only for a qualitative comparison between AGB model predictions
and the observed abundances. However, the dilution range of 0.5-0.7 
also produces a good match to the observed $s$-process abundance 
distribution (see below), with the exception of the Pb abundance. 
Diluting the C and N abundances by 0.5--0.7 dex gives [C/Fe] = 
$+$0.89--1.09, and [N/Fe] = $-0.24$ to $-0.04$ (with corresponding
values of $A(\mathrm{C}) = 8.21-8.41$ and $A(\mathrm{N}) = 6.48-6.68$, respectively).
If we assume 
that HD~123396 has already experienced the {\it first dredge-up 
(FDU)} that takes place following core H-exhaustion, then there 
would be further decreases in C followed by an increase 
in N. 

The low N abundance that we find for HD 123396 may therefore be 
explained by dilution, even if the star has gone through the FDU.
Stars with masses less  than $\approx 2{\it M_{\odot}}$ are observed to experience further mixing of Li and CN processed material
after the luminosity bump \citep{gilroy89,charbonnel94,lind09}.
If our star has experienced such mixing then there should be 
corresponding reductions in C and the $^{12}$C/$^{13}$C ratio to $\approx
5-10$, followed by an increase in N. 
The FDU predicts no change to the elemental O abundance,
and the predictions from the AGB model are [O/Fe] $=0.22$ prior
to dilution and [O/Fe] = $-$0.48 to $-$0.28 after. However, due to its low
metallicity the star was probably alpha-enhanced initially, 
resulting in [O/Fe] = 0 after dilution. 

In terms of neutron-capture abundances, the model with 
our standard partially mixed proton mass of  0.002{\it $M_{\odot}$} 
gives a predicted value of [(Y+Zr+Nd)/Fe]/3 = 1.34 dex. 
After dilution of 0.50--0.70 dex this yields 0.64 - 0.84. A
dilution of 0.60 dex matches the \citet{allen06a} data point
reasonably well, and gives [F/O] = 0.72. To match the abundances 
of the observed [$ls$/Fe] and [$hs$/Fe] ratios given by \citet{allen06a}, 
it is necessary to use a lower dilution factor of 0.30 dex. 
On the other hand, a dilution factor greater than +1.0 dex
is required to reduce the high predicted Pb abundance 
of [Pb/Fe] $=2.61$ to wthat is observed (1.20).
Ignoring uncertainties in neutron-capture cross sections, the main 
factors affecting the production of $s$-process elements in 
low-mass AGB stars are the metallicity of the star, the stellar 
mass, and the size of the $^{13}$C pocket 
\citep[see discussion in e.g.,][]{bisterzo10}.
We can experiment with reducing the mass of the proton profile to 
show the effect of one of these uncertainties. We have calculated one 
model where we reduced the mass of the proton profile by a factor 
of 5 (0.0004{\it $M_{\odot}$}) and obtain [(Y+Zr+Nd)/Fe]/3 $=0.73$ 
and [Pb/Fe] = 1.96. Here a dilution factor of 0.5--0.7 dex 
gives [(Y+Zr+Nd)/Fe]/3 between 0.03--0.23, and [Pb/Fe] abundances 
between 1.26--1.46. While it is possible to match the observed Pb 
abundance with the model with a reduced $^{13}$C pocket,
it is now no longer possible to match the lighter neutron-capture
abundances. 

It may well be possible to fine tune other model parameters,
such as stellar mass, composition, or $^{13}$C pocket size, in order to obtain
a better fit to the observed abundance 
distribution. For example, \citet{husti09} compare their AGB model 
predictions to $s$-process barium star abundances.
They were able to provide a reasonable match to the composition
of HD~123396 by assuming a 1.5{\it $M_{\odot}$} star of $Z=0.001$ and
using a $^{13}$C pocket that was a factor of 4.5 times less than 
their standard, along with a dilution factor of 1.0 dex. 
Given the many uncertainties inherent in current AGB models 
including mass loss, convection, and the formation of $^{13}$C 
pockets \citep[we refer to][for a discussion]{herwig05}, 
we conclude that the AGB mass transfer scenario is shown to be 
quantitatively viable in explaining the light and heavy 
elements in this object.

\section{Concluding remarks}

Employing high-resolution and high S/N data acquired with the Gemini
Observatory Phoenix spectrograph, we have derived CNO, F, Na, and Fe abundances
in the metal-deficient barium red giant star HD~123396. F along with C, N,
and s-process elements are important for constraining the formation processes
of chemically peculiar objects.

The detailed chemical abundance analysis performed here indicates that the nucleosynthetic
history of HD~123396 is explained well in the context of the AGB mass
transfer hypothesis. Furthermore, we
derive, for the first time, [F/O] of a field star (despite its chemical
peculiarity) in the regime spanned by the GC data ([Fe/H]
$\approx$ $-$1). Our results confirm that the GC and field stars have 
different values of [F/O] at a given $A(\mathrm{O})$ abundance. However, gathering more
data is vitally important in order not only to draw firm conclusions on the
nucleosynthetic origin of F in the Galaxy, but also to investigate 
how chemically peculiar objects are formed and how the different 
stellar populations (field versus GCs) are chemically related.

\begin{acknowledgements}
AAB acknowledges CNPq (PDE, 200227/2008-4) and FONDECYT (3100013) for financial
support and Dr. S. Ryder, from the Australian Gemini Office, for his
assistance during the Phase II definitions of our Gemini/Phoenix observations.
JM acknowledges support from FAPESP 
(2010/50930-6), USP (Novos Docentes), and CNPq (Bolsa de produtividade).
We are grateful to Drs. D. Allen and R. Gallino for kindly answering our
questions on their results and to Dr. M. Lugaro for help in preparing the
nuclear network used in the nucleosynthesis code. 
We are also grateful to an anonymous referee
for detailed and useful comments on a previous version of the paper. 
The spectra were obtained as part of the programme GS-2009A-Q-26.
Based on observations obtained at the Gemini Observatory, which is operated by the 
Association of Universities for Research in Astronomy, Inc., under a cooperative agreement 
with the NSF on behalf of the Gemini partnership: the National Science Foundation (United 
States), the Science and Technology Facilities Council (United Kingdom), the 
National Research Council (Canada), CONICYT (Chile), the Australian Research Council (Australia), 
Minist\'{e}rio da Ci\^{e}ncia e Tecnologia (Brazil), 
and Ministerio de Ciencia, Tecnolog\'{i}a e Innovaci\'{o}n Productiva
(Argentina).

\end{acknowledgements}

\bibliography{mnemonic,/home/akarakas/biblio/library}

\end{document}